\documentclass[12pt,letterpaper]{article}
\input{epsf}
\usepackage[pdftex]{graphicx,color}
\input{psfig}
\usepackage{amsmath,amssymb}
\usepackage{fancyhdr}
\newcommand\ltap{\
  \raise.3ex\hbox{$<$\kern-.75em\lower1ex\hbox{$\sim$}}\ }
\newcommand\gtap{\
  \raise.3ex\hbox{$>$\kern-.75em\lower1ex\hbox{$\sim$}}\ }

\newcommand\simge{\mathrel{%
   \rlap{\raise 0.511ex \hbox{$>$}}{\lower 0.511ex \hbox{$\sim$}}}}
\newcommand\simle{\mathrel{
   \rlap{\raise 0.511ex \hbox{$<$}}{\lower 0.511ex \hbox{$\sim$}}}}

\newcommand{\slashchar}[1]%
        {\kern .25em\raise.18ex\hbox{$/$}\kern-.70em #1}
\def\lsim{\mathrel{\raise.3ex\hbox{$<$\kern-.75em\lower1ex\hbox{$\sim$}}}}
\def\gsim{\mathrel{\raise.3ex\hbox{$>$\kern-.75em\lower1ex\hbox{$\sim$}}}}

\newcommand\CH{{\cal H}}

\newcommand\CM{{\cal M}}

\newcommand\CO{{\cal O}}

\newcommand\CZ{{\cal Z}}

\newcommand\be{\begin{equation}}
\newcommand\ee{\end{equation}}
\newcommand\bea{\begin{eqnarray}}
\newcommand\eea{\end{eqnarray}}
\newcommand\ba{\begin{array}}
\newcommand\ea{\end{array}}
\newcommand\nn{\nonumber}

\newcommand{\fourth}{\ensuremath{\frac{1}{4}}}

\newcommand{\thalf}{\textstyle{\frac{1}{2}}}

\newcommand{\tfourth}{\textstyle{\frac{1}{4}}}

\newcommand\dagg{\dagger}

\newcommand\gev{{\rm GeV}}
\newcommand\tev{{\rm TeV}}

\newcommand\pb{{\rm pb}}

\newcommand\fb{{\rm fb}}
\newcommand\ifb{{\rm fb}^{-1}}

\newcommand\etmiss{\slashchar{E}_T}

\newcommand\cbeta{c_\beta}
\newcommand\sbeta{s_\beta}
\newcommand\cbetap{c_{\beta'}}
\newcommand\sbetap{s_{\beta'}}
\newcommand\cdelta{c_\delta}
\newcommand\sdelta{s_\delta}

\newcommand{\LGW}{\Lambda_{\rm GW}}

\begin{document}

\title{
\vspace{-50mm}
{\Large{\bf Phenomenology of the new light Higgs bosons in
    Gildener-Weinberg~models}}\\
} \author{ {\large Kenneth Lane$^1$\thanks{lane@bu.edu}} \, and\, Eric
  Pilon$^2$\thanks{pilon@lapth.cnrs.fr}\\
  {\large $^1$Department of Physics, Boston University}\\
  {\large 590 Commonwealth Avenue, Boston, Massachusetts 02215, USA}\\
  {\large Laboratoire d'Annecy-le-Vieux de Physique Th\'eorique}\\
{\large UMR5108 , Universit\'e de Savoie, CNRS}\\
  {\large B.P. 110, F-74941, Annecy-le-Vieux Cedex, France}\\
} \maketitle

\vspace{-1.0cm}

\begin{abstract}

  Gildener-Weinberg (GW) models of electroweak symmetry breaking are
  especially interesting because the low mass and nearly Standard Model
  couplings of the $125\,\gev$ Higgs boson, $H$, are protected by approximate
  scale symmetry. Another important but so far under-appreciated feature of
  these models is that a sum rule bounds the masses of the new charged and
  neutral Higgs bosons appearing in {\em all} these models to be below about
  $500\,\gev$. Therefore, they are within reach of LHC data currently or soon
  to be in hand. Also so far unnoticed of these models, certain cubic and
  quartic Higgs scalar couplings vanish at the classical level. This is due
  to spontaneous breaking of the scale symmetry. These couplings become
  nonzero from explicit scale breaking in the Coleman-Weinberg loop expansion
  of the effective potential. In a two-Higgs doublet GW model, we calculate
  $\lambda_{HHH} \simeq 2(\lambda_{HHH})_{\rm SM} = 64\,\gev$. This
  corresponds to $\sigma(pp \to HH) \cong 15$--$20\,\fb$, its {\em minimum}
  value for $\sqrt{s} = 13$--$14\,\tev$ at the LHC. It will require at least
  the $27\,\tev$ HE-LHC to observe this cross section. We also find
  $\lambda_{HHHH} \simeq 4(\lambda_{HHHH})_{\rm SM} = 0.129$, whose
  observation in $pp \to HHH$ requires a $100\,\tev$~collider. Because of the
  above-mentioned sum rule, these results apply to {\em all} GW models. In
  view of this unpromising forecast, we stress that LHC searches for the new
  relatively light Higgs bosons of GW models are by far the surest way to
  test them in this decade.

  \end{abstract}



\section*{I. Synopsis}

Section~II of this paper reviews the Gildener-Weinberg (GW) mechanism for
producing a model of a naturally light and aligned Higgs boson, $H$, in
multi-Higgs-scalar models of electroweak symmetry
breaking~\cite{Gildener:1976ih}. This is done in the context of a two-Higgs
doublet model (2HDM) due to Lee and Pilaftsis~\cite{Lee:2012jn}. The
tree-level Higgs potential in GW models is scale-invariant, but that symmetry
can be spontaneously broken, resulting in $H$ as a massless dilaton with
exactly Standard Model (SM) couplings to gauge bosons and fermions. This
scale symmetry is explicitly broken in one-loop order of the Coleman-Weinberg
effective potential~\cite{Coleman:1973jx}, resulting in $M_H^2 > 0$, but only
small deviations from its exact SM couplings. An important corollary of the
formula for $M_H^2$ is a sum rule for the masses of the additional Higgs
scalars, generically $\CH$. In {\em any} GW model of electroweak
breaking in which the only weak bosons are $W^\pm$ and $Z^0$ and the only
heavy fermion is the top quark, the sum rule in first-order loop-perturbation
theory is~\cite{Lee:2012jn,Hashino:2015nxa,Lane:2018ycs}
\be\label{eq:gensum}
\left(\sum_{\CH} M_{\CH}^4\right)^{1/4} = 540\,\gev.
\ee
In the GW-2HDM model, the additional Higgs bosons are a charged pair,
$H^\pm$, and one $C\!P$-even and one $C\!P$-odd scalar, which we call $H_2$ and $A$.
This sum rule has profound consequences for the phenomenology of GW models
that this paper emphasizes. For example, in a search for these new Higgses,
care must be taken in using the sum rule to estimate the light scalar's mass
when the other scalar masses are assumed to exceed 400--500~GeV.

In Sec.~III we discuss features of the cubic and quartic Higgs boson
self-couplings peculiar to GW models. As a consequence of unbroken scale
invariance in the classical Higgs potential, certain of them vanish. These
couplings do become nonzero once the scale symmetry is explicitly broken. We
calculate the most important of these, finding that the experimentally most
relevant ones, $\lambda_{HHH}$ and $\lambda_{HHHH}$, imply $\sigma(pp \to HH)$
and $\sigma(pp \to HHH)$ too small to detect at even the
High-Luminosity (HL) LHC~\cite{Agrawal:2019bpm}. Again, because of the sum
rule~(\ref{eq:gensum}), this conclusion is true in {\em all} GW models of
electroweak symmetry breaking, regardless of their Higgs sector.

This leads to Sec.~IV where we refocus on direct searches at the LHC for the
new light Higgs bosons of GW models. We briefly summarize these Higgses' main
search channels and the status of these searches. Substantial progress is in
reach of data in hand or to be collected in the near future. There is nothing
exotic about these searches; what is required for discovery or exclusion is
greater sensitivity at relatively low masses.

\section*{II. The Two-Higgs Doublet Model}

In 1976, E.~Gildener and S.~Weinberg (GW) proposed a scheme, based on broken
scale symmetry, to generate a light Higgs boson in multi-scalar models of
electroweak symmetry breaking. In essence, their motivation was to generalize
the work of S.~Coleman and E.~Weinberg~\cite{Coleman:1973jx} to completely
general electroweak models, with arbitrary gauge groups and representations
of the fermions and scalars. What GW did not appreciate then --- there was no
reason for them to --- was that their Higgs boson was also
aligned~\cite{Gunion:2002zf}. That is, of all the scalars, its couplings to
gauge bosons and fermions were exactly those of the single Higgs boson of the
Standard Model (SM)~\cite{Weinberg:1967tq}. Like the Higgs boson's mass, its
alignment is protected by the approximate scale symmetry~\cite{Lane:2018ycs}.

GW assumed an electroweak Lagrangian whose Higgs potential $V_0$ has only
quartic interactions. With no quadratic nor cubic Higgs couplings and,
assuming that gauge boson and fermion masses arise entirely from their
couplings to Higgs scalars, the GW theory is scale invariant at the classical
level. This Lagrangian may, however, have a nontrivial extremum. If it does,
it is along a ray in scalar-field space and it is a flat minimum if the
quartic couplings satisfy certain positivity conditions. Thus, scale symmetry
is spontaneously broken at tree level, and there is a massless (Goldstone)
dilaton, $H$, which GW called the ``scalon''. Higgs alignment is a simple
consequence of the linear combination of fields composing $H$ having the {\em
  same form} as the Goldstone bosons $w^\pm$ and $z$ that become the
longitudinal components of the $W^\pm$ and $Z$ bosons; see
Eqs.~(\ref{eq:mevec}) below.

Importantly, scale symmetry is explicitly broken by the first-order term
$V_1$ in the Coleman-Weinberg loop expansion of the effective scalar
potential~\cite{Coleman:1973jx}: $V_0 + V_1$ can have a deeper minimum than
the trivial one at zero fields. If it does, it occurs at a specific vacuum
expectation value (VEV) $\langle H\rangle = v$, explicitly breaking scale
invariance. Then $M_H$ and all other masses in the theory are proportional
to~$v$. The GW scheme is the only one we know in which the entire breaking of
scale and electroweak symmetries is caused by the same electroweak operator,
namely, $\langle H\rangle$. Hence, the dilaton decay constant
$f = v$~\cite{Bellazzini:2012vz}, which we take to be $246\,\gev$.

In 2012, Lee and Pilaftsis (LP) proposed a simple 2HDM model of the GW
mechanism employing the Higgs doublets~\cite{Lee:2012jn}:
\be\label{eq:Phii}
\Phi_i = \frac{1}{\sqrt{2}}\left(\ba{c}\sqrt{2} \phi_i^+ \\ \rho_i + i
  a_i \ea\right), \quad i = 1,2.
\ee
Here, $\rho_i$ and $a_i$ are neutral $C\!P$-even and odd fields. Their potential
is
\bea\label{eq:Vzero}
V_0(\Phi_1,\Phi_2) &=&\lambda_1 (\Phi_1^\dagg \Phi_1)^2 +
\lambda_2 (\Phi_2^\dagg \Phi_2)^2 +
\lambda_3(\Phi_1^\dagg \Phi_1)(\Phi_2^\dagg \Phi_2)\nn \\
&+& \lambda_4(\Phi_1^\dagg \Phi_2)(\Phi_2^\dagg \Phi_1)+
\thalf\lambda_5\left((\Phi_1^\dagg \Phi_2)^2 + (\Phi_2^\dagg
  \Phi_1)^2\right).
\eea
All five quartic couplings are real so that $V_0$ is $C\!P$-invariant as
well. This potential is consistent with a $\CZ_2$ symmetry that prevents
tree-level flavor-changing interactions among fermions, $\psi$, induced by
neutral scalar exchange~\cite{Glashow:1976nt}:
\be\label{eq:Z2}
\Phi_1 \to -\Phi_1,\,\, \Phi_2 \to \Phi_2, \quad
\psi_L \to -\psi_L,\,\, \psi_{uR} \to \psi_{uR},\,\,
\psi_{dR} \to \psi_{dR}.
\ee
This is the usual type-I 2HDM~\cite{Branco:2011iw}, but with $\Phi_1$ and
$\Phi_2$ interchanged; we refer henceforth to this version of the model as
the GW-2HDM. This choice of Higgs couplings differs from LP's choice of
type-II~\cite{Lee:2012jn}. It was made to remain consistent with limits from
CMS~\cite{Khachatryan:2015qxa} and ATLAS~\cite{Aaboud:2018cwk} on charged
Higgs decay into $t\bar b$. The limits from these papers are consistent with
$\tan\beta \simle 0.5$ for $M_{H^\pm} \simle 500\,\gev$. This range of
$\tan\beta$ also suppresses $gg \to A(H') \to \bar bb,\,\bar tt$, where $A(H')$
is a $C\!P$-odd (even) Higgs, relative to a heavy Higgs boson~$H$ with SM
couplings. See the discussion and references in Ref.~\cite{Lane:2018ycs}.

The potential $V_0$ can have a flat minimum along the ray
\be\label{eq:theray}
\Phi_{1\beta} = \frac{1}{\sqrt{2}} \left(\ba{c} 0\\ \phi\,\cbeta
  \ea\right),\quad
\Phi_{2\beta} = \frac{1}{\sqrt{2}} \left(\ba{c} 0\\ \phi\,\sbeta \ea\right).
\ee
Here $\phi > 0$ is any real mass scale, $\cbeta = \cos\beta$ and
$\sbeta = \sin\beta$. The nontrivial tree-level extremal conditions are (for
$\beta \neq 0, \pi/2$):
\be\label{eq:first}
\lambda_1\cbeta^2 + \thalf\lambda_{345}\sbeta^2 = 0,\qquad
\lambda_2\sbeta^2 + \thalf\lambda_{345}\cbeta^2 = 0,
\ee
where $\lambda_{345} = \lambda_3 + \lambda_4 + \lambda_5$. Scale symmetry is
spontaneously, but not yet explicitly, broken. Note that
$V_{0\beta} = V_0(\Phi_{1\beta},\Phi_{2\beta}) = 0$, degenerate with the
trivial vacuum. The squared ``mass'' matrices of the $C\!P$-odd, charged, and
$C\!P$-even scalars are given by
\be
\label{eq:Mmsq}
\CM^2_S = -\lambda_S \phi^2 \left(\ba{cc} \sbeta^2 & -\sbeta
  \cbeta\ \\ -\sbeta \cbeta& \cbeta^2 \ea\right), \\
\ee
where the subscript $S = H_{0^-}$, $H^\pm$, and $H_{0^+}$ In terms of the
quartic couplings in the Higgs potential, they are
$\lambda_{H_{0^-}} = \lambda_5$;
$\lambda_{H^\pm} = \thalf(\lambda_4 + \lambda_5) = \thalf \lambda_{45}$; and
$\lambda_{H_{0^+}} = \lambda_{345}$. All $\lambda_S$ are negative to ensure
non-negative eigenvalues of the matrices. The respective eigenvectors and
eigenvalues are:
\bea\label{eq:mevec}
\left(\ba{c} z \\ A\ea\right) &=& \left(\ba{cc} \cbeta & \sbeta\\
     -\sbeta & \cbeta\ea\right) \left(\ba{c} a_1 \\ a_2\ea\right),
\quad M_z^2 = 0, \,\,\, M_A^2  = -\lambda_5 \phi^2;\nn\\
%
\left(\ba{c} w^\pm \\ H^\pm\ea\right) &=& \left(\ba{cc} \cbeta & \sbeta\\
     -\sbeta & \cbeta\ea\right) \left(\ba{c} \phi_1^\pm \\ \phi_2^\pm\ea\right),
\quad M_{w^\pm}^2 = 0, \,\,\, M_{H^\pm}^2 = -\thalf\lambda_{45}
\phi^2;\nn \\
%
\left(\ba{c} H \\ H'\ea\right) &=& \left(\ba{cc} \cbeta & \sbeta\\
     -\sbeta & \cbeta\ea\right) \left(\ba{c} \rho_1 \\ \rho_2\ea\right),
\quad M_H^2 = 0, \,\,\, M^2_{H'}  = -\lambda_{345} \phi^2.
\eea

The one-loop effective potential, presented in Ref.~\cite{Lee:2012jn}, is
given by
\bea\label{eq:Vone}
V_1 &=&\frac{1}{64\pi^2}
          \biggl[6M_W^4\left(-\frac{5}{6} + \ln\frac{M_W^2}{\LGW^2}\right)
          + 3M_Z^4\left(-\frac{5}{6} + \ln\frac{M_Z^2}{\LGW^2}\right) \nn\\
          &&\quad+ M_{H'}^4\left(-\frac{3}{2}+\ln\frac{M_{H'}^2}{\LGW^2}\right) +
          M_A^4\left(-\frac{3}{2} +\ln\frac{M_A^2}{\LGW^2}\right) \nn\\
          &&\quad +
          2M_{H^\pm}^4\left(-\frac{3}{2}+\ln\frac{M_{H^\pm}^2}{\LGW^2}\right)
          -12m_t^4\left(-1 + \ln\frac{m_t^2}{\LGW^2}\right)\biggr],
\eea
where $\Lambda_{GW}$ is the GW renormalization scale (related to the Higgs
VEV~$v$ by Eq.~(40,41) in LP). The background field-dependent masses in
$V_1$ are
\bea\label{eq:bgmasses}
M_W^2 &=& \thalf g^2 \left(\Phi_1^\dagg \Phi_1 + \Phi_2^\dagg \Phi_2\right),
\nn\\
M_Z^2 &=& \thalf(g^2+g^{\prime\, 2}) \left(\Phi_1^\dagg \Phi_1 + \Phi_2^\dagg
\Phi_2\right), \nn\\
M_A^2 &=& -2\lambda_5\left(\Phi_1^\dagg \Phi_1 + \Phi_2^\dagg
  \Phi_2\right), \nn\\
M_{H^\pm}^2 &=& -\lambda_{45}\left(\Phi_1^\dagg \Phi_1 + \Phi_2^\dagg
  \Phi_2\right), \nn\\
M_{H'}^2 &=& -2\lambda_{345}\left(\Phi_1^\dagg \Phi_1 + \Phi_2^\dagg
  \Phi_2\right), \nn\\
m_t^2 &=& \Gamma_t^2\, \Phi_1^\dagg \Phi_1,
\eea
where $g,g'$ are the electroweak $SU(2)$ and $U(1)$ gauge couplings and
$\Gamma_t =\sqrt{2}m_t/v_1 = \sqrt{2}m_t/v\cos\beta$ is the Higgs-Yukawa
coupling of the top quark. In Eqs.~(\ref{eq:bgmasses}), the $C\!P$-even part of
$\Phi_i$ is the shifted field $v_i + \rho_i$.

The nontrivial extremal conditions for $V_0 + V_1$ are~\cite{Lee:2012jn}
\bea
\label{eq:extrx}
\left.\frac{\partial(V_0+V_1)}{\partial\rho_1}\right\vert_{\langle\,\rangle}
&\propto& \lambda_1\cbeta^2 + \thalf\lambda_{345}\sbeta^2 + \Delta\widehat
    t_1/64\pi^2 = 0,\nn\\
\left.\frac{\partial(V_0+V_1)}{\partial\rho_2}\right\vert_{\langle\,\rangle}
&\propto&  \lambda_2\sbeta^2 + \thalf\lambda_{345}\cbeta^2 + \Delta\widehat
    t_2/64\pi^2 = 0,
\eea
where $\langle\,\rangle$ means that the derivatives of $V_0 + V_1$ are
evaluated at the vacuum expectation values of the fields, and
\bea\label{eq:Deltat}
\Delta\widehat t_i &=&
\frac{4}{v^4}\biggl[2M_W^4\left(3\ln{\frac{M_W^2}{\LGW^2}} - 1\right)
 + M_Z^4\left(3\ln{\frac{M_Z^2}{\LGW^2}} - 1\right)
 + M_{H'}^4\left(\ln{\frac{M_{H'}^2}{\LGW^2}} - 1\right) \nn\\
&\quad& + M_A^4 \left(\ln{\frac{M_A^2}{\LGW^2}} - 1\right)
        + 2M_{H^\pm}^4 \left(\ln{\frac{M_{H^\pm}^2}{\LGW^2}} - 1\right)
        - 12m_t^4 \left(\ln{\frac{m_t^2}{\LGW^2}} - \frac{1}{2}\right)\delta_{i1}
\biggr]_{\langle\,\rangle}.\hspace{1.0cm}
\eea
For nontrivial extrema with $\beta \neq 0,\,\pi/2$, these conditions lead to
a deeper minimum than the zeroth-order ones,
$(V_0 + V_1)_{\rm min} < V_{0\beta} = V_0(0) + V_1(0) = 0$. This minimum
occurs at a particular value~$v$ of the scale $\phi$ which, as we've said, is
identified as the electroweak breaking scale, ~$v = 246\,\gev$. The VEVs of
$\Phi_1$ and $\Phi_2$ are $v_1 = v\cos\beta$ and $v_2 = v\sin\beta$, with
$\tan\beta = v_2/v_1$ as usual in 2HDM.

\begin{figure}[ht!]
 \begin{center}
\includegraphics[width=2.65in, height=2.65in]{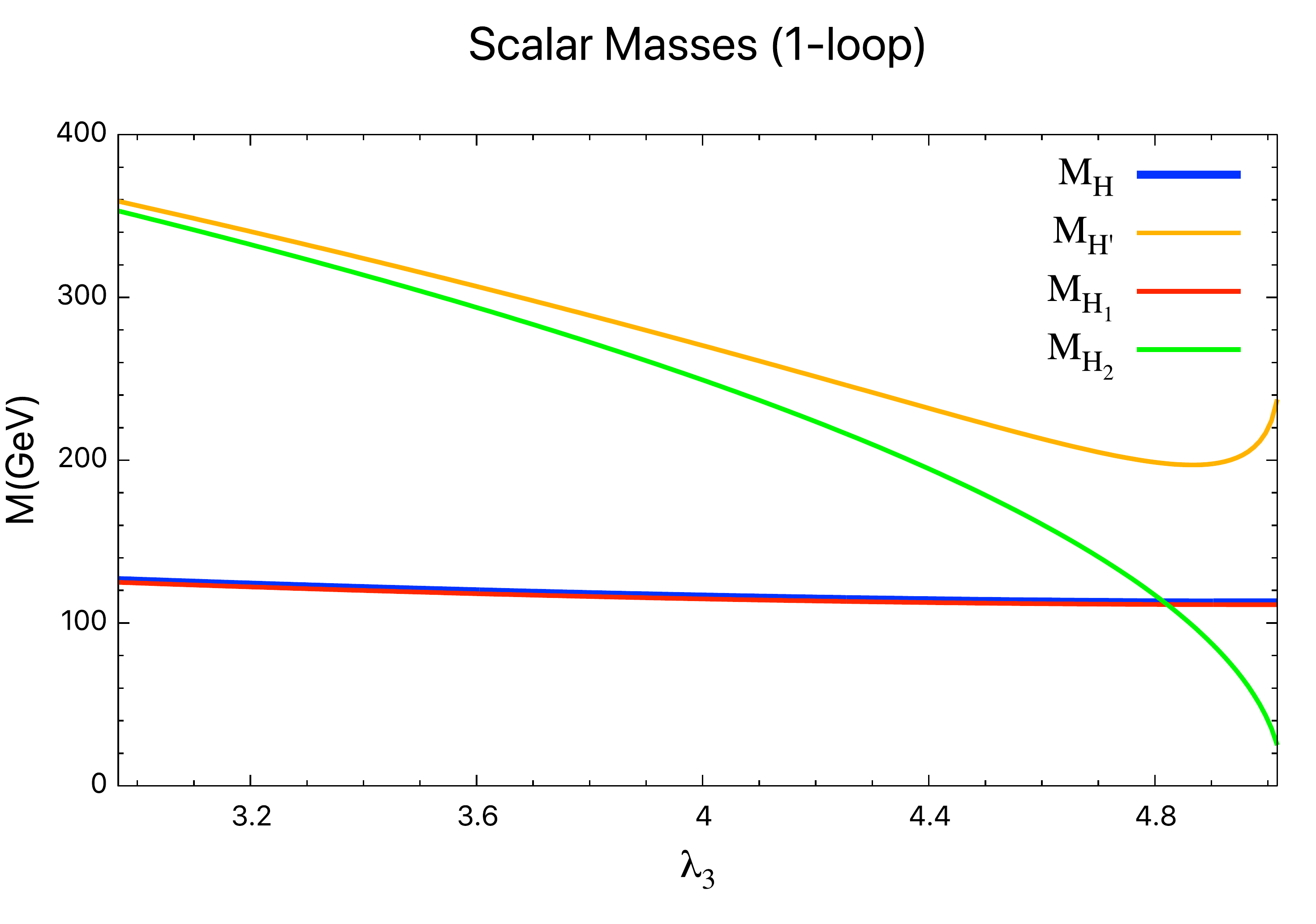}
\includegraphics[width=2.65in, height=2.65in]{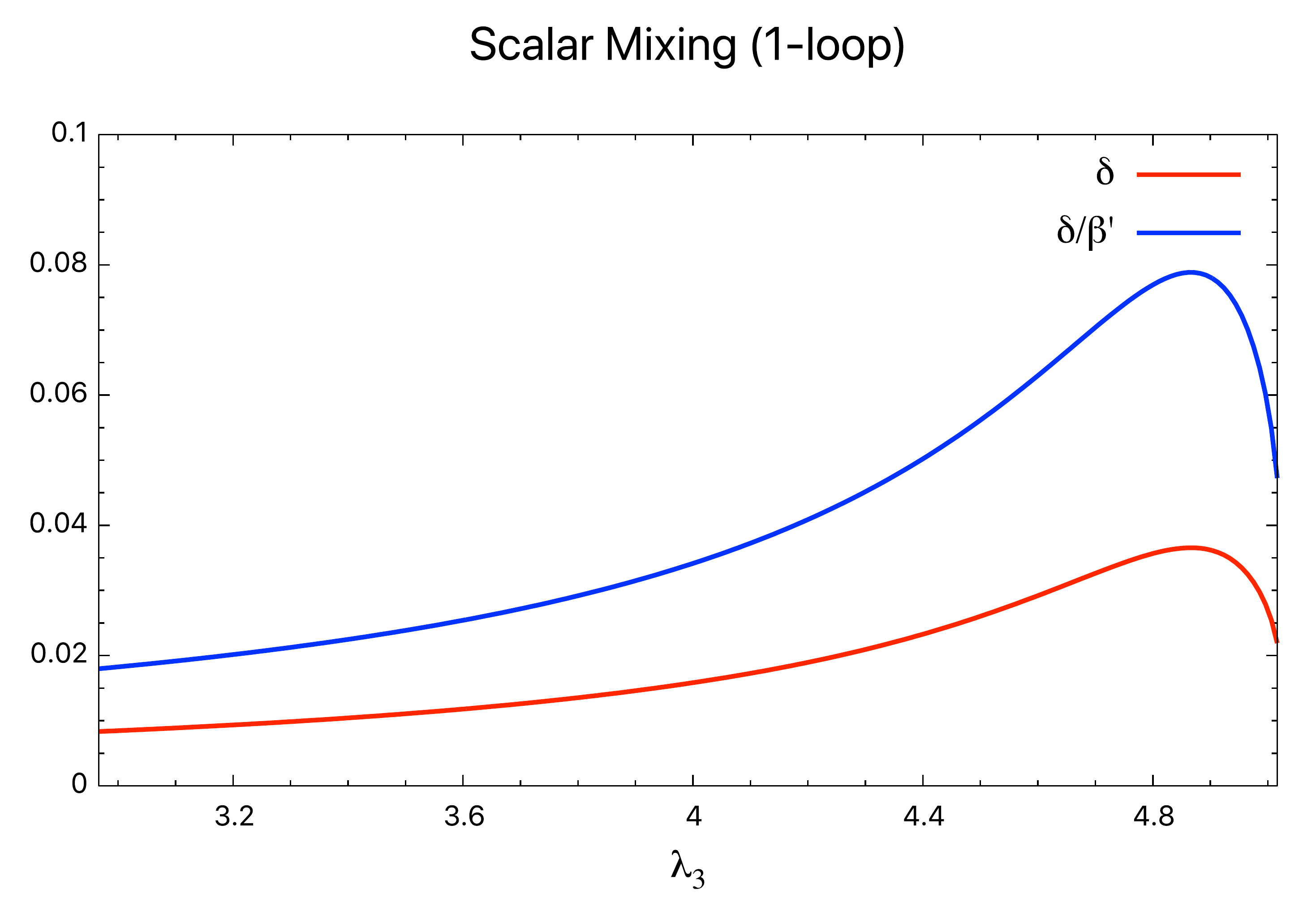}
\caption{Left: The $C\!P$-even Higgs one-loop mass eigenvalues $M_{H_1}$ and
  $M_{H_2}$, the tree-level mass $M_{H'} = \sqrt{-\lambda_{345}}\,v$ and the
  one-loop mass $M_H$ from Eq.~(\ref{eq:MHsq}) as functions of
  $\lambda_3 = (2M_{H^\pm}^2 - M_{H'}^2)/v^2$. Here, $\tan\beta = 0.50$ and
  $M_{H^\pm} = M_A = 390\,\gev$ corresponding to
  $\lambda_4 = \lambda_5 = -2.513$. The input $H \cong H_1$ mass is
  $M_H = 125.0\,\gev$, the corresponding initial $M_{H'} = 353\,\gev$ and
  $\lambda_3 = 2.966$. $M_{H'}$ vanishes at
  $\lambda_3 = 2M^2_{H^\pm}/v^2 = 5.027$. Right: The angle
  $\delta = \beta - \beta'$ measuring the deviation from perfect alignment of
  $H_1$ and the ratio $\delta/\beta$ for $\beta = 0.4637$. The procedure used
  in creating these figures is spelled out in the Appendix of
  Ref.~\cite{Lane:2018ycs}}.
  \label{fig:oneloop}
 \end{center}
 \end{figure}

 The $C\!P$-odd and charged Higgs bosons' masses receive no contribution from
 $V_1$ and, so, they are given by Eqs.~(\ref{eq:mevec}) with $\phi = v$. The
 $C\!P$-even masses, however, receive important corrections from $V_1$. The
 eigenvectors $H_1$ and $H_2$ are
\bea\label{eq:H1H2}
H_1 &=& \cdelta H - \sdelta H' = \cbetap\rho_1 + \sbetap\rho_2,\nn\\
H_2 &=& \sdelta H + \cdelta H' = -\sbetap\rho_1 + \cbetap\rho_2,
\eea
where $\beta' = \beta -\delta$, $\cbetap = \cos\beta'$, etc. The
angle~$\delta$ measures the departure of the Higgs boson $H_1$ from perfect
alignment, and it should be small. Furthermore, the accuracy of first-order
perturbation theory requires $|\delta/\beta| \ll 1$. Both these criteria are
met in calculations with a wide range of input parameters; they are
illustrated in Fig.~\ref{fig:oneloop}. From now on we refer interchangeably
to the 125~GeV Higgs boson as $H_1$ or $H$, as clarity requires. Its mass is
given by~\cite{Gildener:1976ih},\cite{Lee:2012jn},\cite{Lane:2018ycs}
\be\label{eq:MHsq}
M_{H_1}^2 \cong M_H^2 = \frac{1}{8\pi^2 v^2}\left(6M_W^4 + 3M_Z^4
  + M_{H'}^4 + M_A^4 + 2M_{H^\pm}^4 - 12m_t^4\right).
\ee
In accord with first-order perturbation theory, all the masses on the right
side of this formula are obtained from zeroth-order perturbation theory,
i.e., from $V_0$ plus gauge and Yukawa interactions, with $\phi = v$. As we
see in Fig.~\ref{fig:oneloop}, the Higgs masses $M_H$ and $M_{H_1}$ derived
from Eq.~(\ref{eq:MHsq}) and from diagonalizing the one-loop mass
matrix $\CM_{H_{0^+}}$ respectively, are extremely close, as they should be.

\begin{figure}[ht!]
\includegraphics[width=1.0\textwidth]{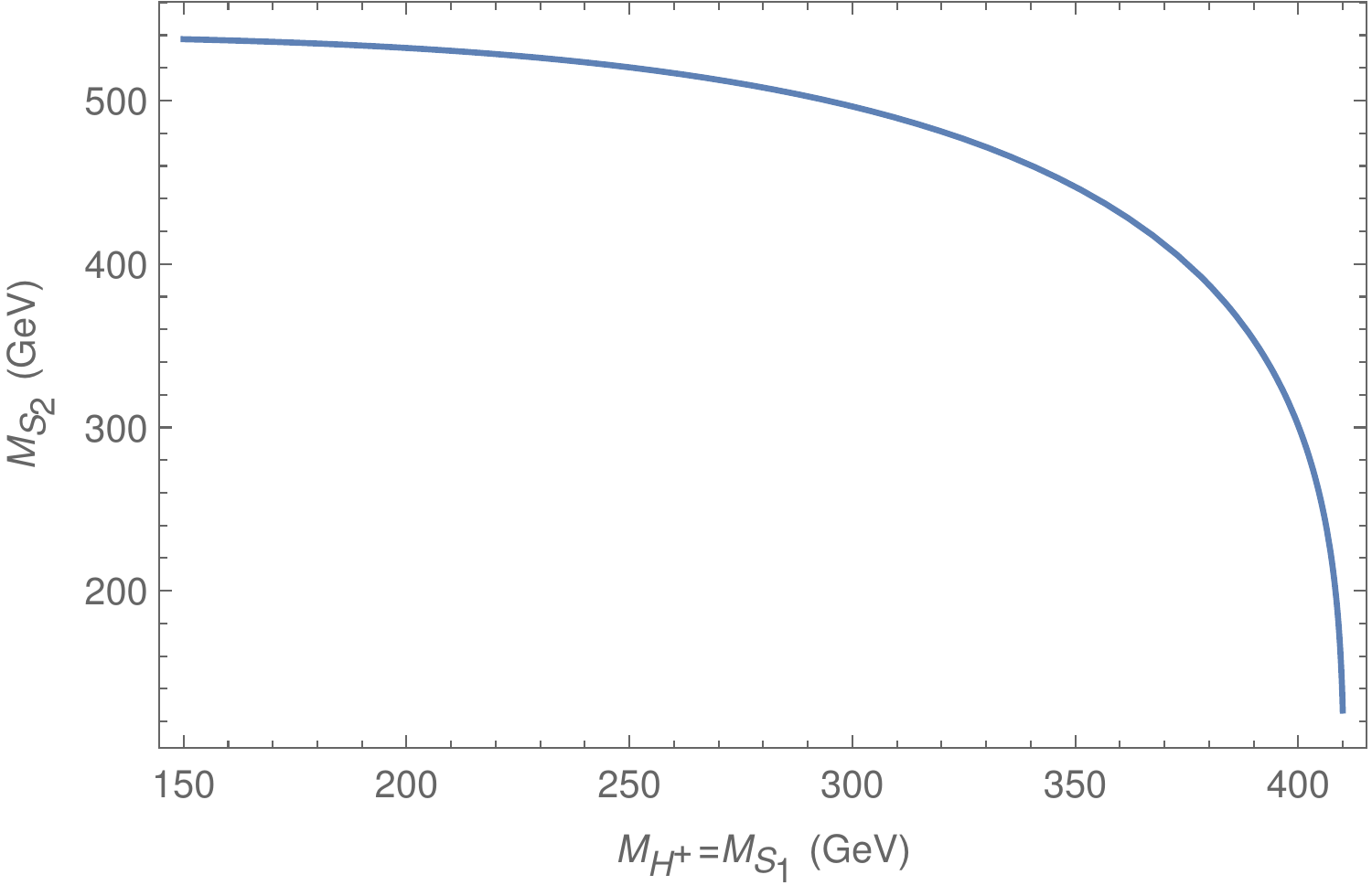}
\caption{The mass of the neutral Higgs $S_2 = H'$ $(S_2 = M_A)$ as a function
  of the common mass of $H^\pm$ and the other neutral Higgs, $S_1 = A$
  $(S_1 = H')$, from Eqs.(\ref{eq:MHsq},\ref{eq:MHsum}) with
  $M_H = 125\,\gev$. Note the considerable sensitivity of $M_{S_2}$ to small
  changes in $M_{H^+} = M_{S_1}$ when it is large. From
  Ref.~\cite{Lane:2018ycs}}
\label{fig:mlone}
\end{figure}

This formula can be used in two related ways. First, assuming that there are
no other heavy fermions and weak bosons, it implies a sum rule on all the new
scalar masses in this GW-2HDM~\cite{Lee:2012jn,Hashino:2015nxa,Lane:2018ycs}:
\be\label{eq:MHsum}
\left(M_{H'}^4 + M_A^4 + 2M_{H^\pm}^4\right)^{1/4} = 540\,\gev.
\ee
The sum rule is illustrated in Fig.~\ref{fig:mlone} for
$M_{H} \cong M_{H_1} = 125\,{\gev}$ and $M_{H^\pm} = M_{S_1}$, where
$M_{S_1} = M_A$ or $M_{H'}$; the mass of the other neutral scalar, $M_{S_2}$,
is plotted against~$M_{H^\pm} = M_{S_1}$. The smallness of~$\delta$ in
Fig.~\ref{fig:oneloop} and the magnitude of Higgs couplings we obtain in
Sec.~III give us confidence that the one-loop approximation~(\ref{eq:MHsq}) is
reliable. Still, we would not be surprised if higher-order corrections change
the right side of Eq.~(\ref{eq:MHsum}) by $\CO(100\,\gev)$. The important
point is that the sum rule tells us that new Higgs bosons should be found at
surprisingly low masses. To repeat: this sum rule holds in {\em any}
GW model of electroweak breaking in which the only weak bosons are $W$ and
$Z$ and the only heavy fermion is the top quark. Thus, the larger the Higgs
sector, the lighter will be the masses of at least some of the new Higgs
bosons expected in a GW model.


%
\begin{figure}[ht!]
\includegraphics[width=1.0\textwidth]{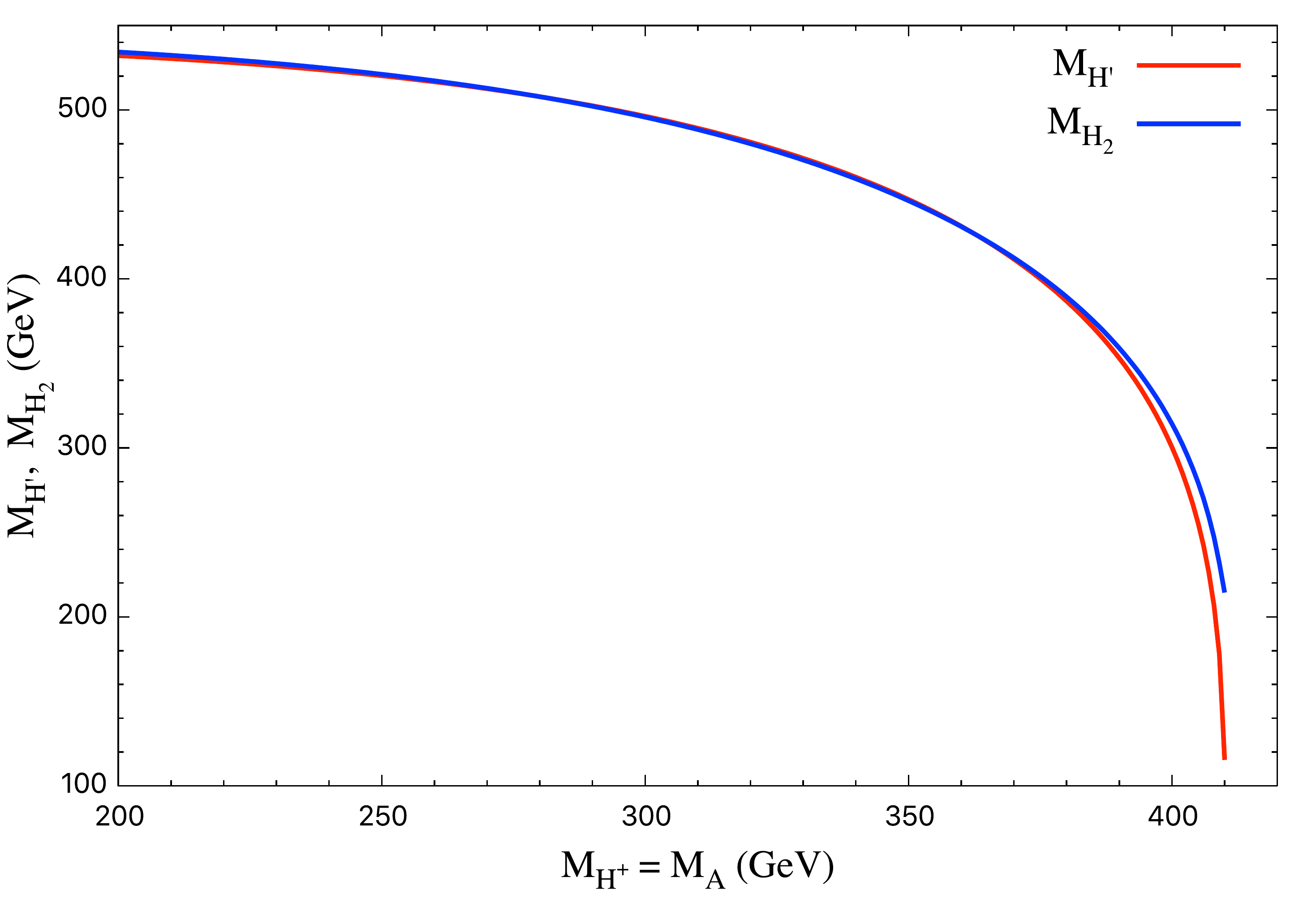}
\caption{The tree-approximation mass $M_{H'}$ of the $C\!P$-even Higgs calculated
  from the sum rule~(\ref{eq:MHsum}) and the larger eigenvalue $M_{H_2}$ of
  the one-loop corrected $C\!P$-even mass matrix $\CM_{H_{0^+}}$. Both are
  calculated as a function of $M_{H^\pm} = M_A$. $M_{H'}$ starts to dive to
  zero at $M_{H^\pm} \cong 370\,\gev$ and becomes zero at
  $M_{H^\pm} \cong 410.22\,\gev$.}
\label{fig:CPeven}
\end{figure}
Second, as an instructive example in the present model, we assume that
$M_{H^\pm} = M_A$ and imagine searching for $H_2 \simeq H'$. (The assumption
$M_{H^\pm} = M_A$ is motivated by the fact that it makes the contribution to
the $T$-parameter from the scalars vanish identically~\cite{Battye:2011jj,
  Pilaftsis:2011ed}.)  Due to the sum-rule constraint in
Eq.~(\ref{eq:MHsum}), the mass of $M_{H'}$ is very sensitive to small changes
in $M_{H^\pm}$ when it is large. Fig.~\ref{fig:mlone} suggests we can use the
sum rule until $M_{H'}$ starts to dive to zero. To be quantitative about
this, Fig.~\ref{fig:CPeven} shows $M_{H'}$ and $M_{H_2}$ as a function of
$M_{H^\pm} = M_A$ over the range allowed by the sum rule.\footnote{The only
  model parameter that enters this calculation is $\tan\beta$;
  Fig.~\ref{fig:CPeven} is practically independent of $\tan\beta$.} The two
masses are very nearly equal up to $M_{H^\pm} = 370\,\gev$. At that mass,
$M_{H'} \cong M_{H_2} = 412\,\gev$ and, beyond it, $M_{H'}$ starts its
dive. The most sensible thing to do, in our opinion, is to use the large
$C\!P$-even mass eigenvalue, $M_{H_2}$, over the entire considered range of
$M_{H^\pm} = M_A$.

We will do this for our estimates of the scalars' production cross sections
and decay branching ratios in Sec.~IV. We recommend this approach for searches
by ATLAS and CMS. For example, in a search involving the three GW Higgs
bosons (say, $pp \to A \to Z H_2$ and $pp \to H^\pm \to W^\pm H_2$, with
$H_2 \to b \bar b$), one could use ellipsoidal search regions in
($M_{H^\pm},M_A, M_{H_2}$)-space roughly consistent with $M_{H^\pm} = M_A$
and the sum rule, and calculate the model's predicted
$\sigma\cdot {\rm BR}$'s accordingly. Therefore, as with $H$ and $H_1$, we
refer henceforth to the heavier $C\!P$-even scalar as $H'$ or $H_2$, as clarity
or the situation requires.

\section*{III. Triple and Quartic Higgs Couplings}

In GW models of electroweak symmetry breaking, the tree-level triple-scalar
couplings involving two or three of the Goldstone bosons $H,z,w^\pm$ vanish,
as do the quartic couplings involving three or four of them. This is unlike
any other multi-Higgs model. The reason for this, of course, must be scale
invariance of the tree-level Lagrangian, in particular, that the potential
$V_0$ contains only quartic couplings. But how does it work? We show how in
this section. Then we calculate at one-loop order the triple-scalar couplings
involving at least one $H \cong H_1$ and the quartic coupling
$\lambda_{H_1H_1H_1H_1}$.

The way to see simply why certain scalar couplings vanish is to write $V_0$
in the ``aligned basis'':
\be\label{eq:aligned}
\Phi = \Phi_1\cbeta + \Phi_2\sbeta, \quad \Phi' = -\Phi_1\sbeta +
\Phi_2\cbeta.
\ee
On the ray Eq.~(\ref{eq:theray}) on which $V_0$ has nontrivial extrema, these
fields are
\be\label{eq:alignedray}
\Phi_\beta = \frac{1}{\sqrt{2}} \left(\ba{c} 0\\ \phi \ea\right),\quad
\Phi'_\beta = \frac{1}{\sqrt{2}} \left(\ba{c} 0\\ 0 \ea\right),
\ee
where $\phi \in (0,\infty)$ is a constant mass scale. Then, in terms of the
tree-level mass-eigenstate scalars, the fields $\Phi$, $\Phi'$ are
\be\label{eq:fieldsray}
\Phi = \frac{1}{\sqrt{2}} \left(\ba{c} \sqrt{2}w^+\\ \phi + H + iz\ea\right),\quad
\Phi' = \frac{1}{\sqrt{2}} \left(\ba{c} \sqrt{2}H^+\\ H' + iA \ea\right).
\ee
Rewritten in terms of quartic polynomials in $\Phi$ and $\Phi'$,
Eq.~(\ref{eq:Vzero}) becomes (with $\lambda_{345} = \lambda_3 + \lambda_4 +
\lambda_5$, etc.)
\bea\label{eq:VzeroEP}
V_0 &=& \left[\lambda_1\cbeta^4 + \lambda_2\sbeta^4 +
  \lambda_{345}\sbeta^2\cbeta^2\right] \left(\Phi^\dagg \Phi\right)^2 \nn\\
&&+\left[(2\lambda_2\sbeta^2 +\lambda_{345}\cbeta^2) - (2\lambda_1\cbeta^2
  +\lambda_{345}\sbeta^2) \right]\sbeta\cbeta \left(\Phi^\dagg \Phi\right)
\left(\Phi^\dagg\Phi^\prime+ \Phi^{\prime\,\dagg}\Phi\right) \nn\\
&&+\left[2(\lambda_1+\lambda_2-\lambda_{345})\sbeta^2\cbeta^2 +
  \lambda_3\right] \left(\Phi^\dagg \Phi\right)
                   \left(\Phi^{\prime\,\dagg}\Phi^\prime\right) \nn\\
&&+\left[2(\lambda_1+\lambda_2-\lambda_{345})\sbeta^2\cbeta^2 +
  \lambda_4\right] \left(\Phi^\dagg \Phi^\prime\right)
                   \left(\Phi^{\prime\,\dagg}\Phi\right) \nn\\
&&+\thalf\left[2(\lambda_1+\lambda_2-\lambda_{345})\sbeta^2\cbeta^2 +
    \lambda_5\right] \left[\left(\Phi^\dagg \Phi^\prime\right)^2
    +\left(\Phi^{\prime\,\dagg} \Phi\right)^2\right]\nn\\
&&+\left[(2\lambda_2\cbeta^2 +\lambda_{345}\sbeta^2) - (2\lambda_1\sbeta^2
   +\lambda_{345}\cbeta^2) \right]\sbeta\cbeta \left(\Phi^{\prime\,\dagg}
   \Phi^\prime\right)
 \left(\Phi^\dagg\Phi^\prime+ \Phi^{\prime\,\dagg}\Phi\right) \nn\\
&&+\left[\lambda_1\sbeta^4 + \lambda_2\cbeta^4 +
  \lambda_{345}\sbeta^2\cbeta^2\right] \left(\Phi^{\prime\,\dagg}
  \Phi^\prime\right)^2.
\eea

By virtue of its scale invariance, $V_0$ is a homogeneous polynomial of
degree~four:
\bea\label{eq:Vhpoly}
V_0 &=& {\fourth}\sum_{i=1}^2 \left[\Phi_i^\dagg \frac{\partial V_0}{\partial
    \Phi_i^\dagg} + \frac{\partial V_0}{\partial \Phi_i}\Phi_i\right] \nn \\
&=& {\fourth}\left[\Phi^\dagg \frac{\partial V_0}{\partial
    \Phi^\dagg} + \frac{\partial V_0}{\partial \Phi}\Phi +
                  \Phi^{\prime\,\dagg} \frac{\partial V_0}{\partial
    \Phi^{\prime\,\dagg}} + \frac{\partial V_0}{\partial \Phi'}\Phi' \right].
\eea
Thus, $V_0$ vanishes at {\em any} extremum, in particular for
$\Phi_\beta = (0,\phi)/\sqrt{2}$ and $\Phi'_\beta = (0,0)$, the flat
direction associated with spontaneous scale symmetry breaking. We know that
the conditions for the nontrivial extrema of $V_0$ are those in
Eq.~(\ref{eq:first}). It follows that the coefficients of
$(\Phi^\dagg \Phi)^2$ and
$(\Phi^\dagg \Phi) (\Phi^\dagg \Phi' + \Phi^{'\,\dagg} \Phi)$ terms in $V_0$
vanish. It is easy to see why these coefficients, $C_1$ and $C_2$, had to
vanish. On the ray $\Phi_\beta,\Phi'_\beta$,
\bea \left.\frac{\partial
    V_0}{\partial\Phi}\right\vert_{\Phi_\beta,\Phi'_\beta} &=& 2C_1
 \Phi^\dagg_\beta\left(\Phi^\dagg_\beta \Phi_\beta\right),\\
\left.\frac{\partial V_0}{\partial\Phi'}\right\vert_{\Phi_\beta,\Phi'_\beta}
&=& C_2 \left(\Phi^\dagg_\beta \Phi_\beta\right) \Phi^\dagg_\beta.
\eea
Neither operator vanishes, hence their coefficients must.\footnote{Of course,
  $C_1 = C_2 = 0$ implies the conditions of Eq.~(\ref{eq:first}.} This would
not have happened had $V_0$ also contained polynomials of degree less than
four. That is, spontaneously broken scale invariance is the reason for the
vanishing Goldstone boson couplings at tree level. And it is obvious that
this analysis using homogeneous polynomials of fourth degree generalizes to
{\em any} GW model of the electroweak interactions.

Using the tree-level extremal conditions, the nonzero coefficients in $V_0$
are simplified by using
\bea\label{eq:Cnonzero}
&&2(\lambda_1 + \lambda_2-\lambda_{345})\sbeta^2\cbeta^2 = -\lambda_{345},\\
&&\left[(2\lambda_2\cbeta^2 +\lambda_{345}\sbeta^2) - (2\lambda_1\sbeta^2
   +\lambda_{345}\cbeta^2) \right]\sbeta\cbeta = -2\lambda_{345}\cot 2\beta,\\
&&\lambda_1\sbeta^4 + \lambda_2\cbeta^4+\lambda_{345} \sbeta^2\cbeta^2 = -
2\lambda_{345} \cot^2 2\beta.
\eea
Then,
\bea\label{eq:Vzerofinal}
V_0 &=& -\lambda_{45}\left(\Phi^\dagg\Phi\right)
\left(\Phi^{\prime\,\dagg} \Phi'\right)
-\lambda_{35}\left(\Phi^\dagg\Phi^\prime\right)
                      \left(\Phi^{\prime\,\dagg} \Phi\right)
 -\thalf\lambda_{34}\left[\left(\Phi^\dagg \Phi'\right)^2
    +\left(\Phi^{\prime\,\dagg} \Phi\right)^2\right]\nn\\
  &&-2\lambda_{345}\cot 2\beta \left(\Phi^{\prime\,\dagg} \Phi'\right)
 \left(\Phi^\dagg\Phi'+ \Phi^{\prime\,\dagg}\Phi\right)
-2\lambda_{345}\cot^2 2\beta\left(\Phi^{\prime\,\dagg} \Phi'\right)^2.
\eea
From this, the masses in Eq.~(\ref{eq:Mmsq}) may be read off from the first
three terms.

With foreknowledge, we now put $\phi = v = 246\,\gev$. Then the nonzero cubic
terms terms in the tree-level potential, written in terms of mass eigenstate
scalars of $V_0$, are:\footnote{Of course, the electroweak Goldstone fields
  $w^\pm,z$ are absent in the unitary gauge, but must be retained in
  renormalizable gauges.}
\bea\label{eq:V0cubic}
V_0({\rm cubic}) &=& -\thalf\lambda_{45}\,vH\left[(H')^2 + A^2 +
  2H^+H^-\right]\nn\\
&& -\thalf\lambda_{35}\,v\bigl[H'(HH' + zA + w^+H^- + H^+w^-)\nn\\
&&\qquad + A(HA-zH') +iA(w^+ H^- - H^+w^-)\bigr]\nn\\
&&-\thalf\lambda_{34}\,v\bigl[H'(HH' + zA + w^+H^- + H^+w^-)\nn\\
&&\qquad - A(HA - zH') -iA(w^+H^- - H^+w^-)\bigr]\nn\\
&& -\lambda_{345}\cot 2\beta\,vH'\left[(H')^2 + A^2 + 2H^+H^-\right].
\eea
The quartic terms are:
\bea\label{eq:V0quartic}
V_0({\rm quartic}) &=& -\tfourth\lambda_{45}(H^2+z^2 + 2w^+w^-)\left[(H')^2 +
  A^2 + 2H^+H^-\right]\nn\\
&& -\tfourth\lambda_{35}
\bigl[(HH'+zA+2w^+H^-)(HH'+zA+2H^+w^-)\nn\\
&&\qquad  + (HA - zH')^2 + 2i(HA - zH')(w^+H^- - H^+w^-)\bigr]\nn\\
&& -\tfourth\lambda_{34}
\bigl[\thalf(HH' + zA + 2w^+H^-)^2 + \thalf(HH' + zA + 2H^+w^-)^2\nn\\
&&\qquad - (HA-zH')^2 - 2i(HA-zH')(w^+H^- - H^+w^-)\bigr]\nn\\
&&-\lambda_{345}\cot 2\beta\left[(H')^2+A^2+2H^+H^-\right]
\left[HH' + zA + w^+H^- + H^+w^-\right]\nn\\
&&-\thalf\lambda_{345}\cot^2 2\beta\left[(H')^2 + A^2 + 2H^+H^-\right]^2.
\eea
Recall from Eq.~(\ref{eq:Mmsq}) that $-\lambda_{345} = M_{H'}^2/v^2$,
$-\lambda_{45} = 2M_{H^\pm}^2/v^2$ and $-\lambda_5 = M_A^2/v^2$.

We turn to the one-loop corrections, focusing on the triple-scalar couplings
involving the $125\,\gev$ Higgs boson, $H \cong H_1$, and the quartic
coupling $\lambda_{H_1H_1H_1H_1}$. For brevity, we include only those cubic
couplings of $H_1$ with itself and with $H_2$. The $H_1AA$ and $H_1H^+H^-$
couplings are similar to $H_1H_2H_2$, as may be inferred from the tree-level
cubics in Eq.~(\ref{eq:V0cubic}) and Table~\ref{tab:lambdas} below. There are
two types of one-loop corrections: (i) those to $V_0$ obtained by writing the
zeroth-order $C\!P$-even fields in terms of $H_1$ and $H_2$,
Eqs.~(\ref{eq:H1H2}), and by using the one-loop extremal conditions,
Eqs.~(\ref{eq:extrx}); (ii) those obtained from $V_1$ in Eq.~(\ref{eq:Vone})
by isolating the coefficients of $H^3$, $H^2H'$, etc.

\medskip

\noindent (i)~With $\rho_i$ shifted by~$v_i$, the cubic $C\!P$-even terms in
$V_0$ are:
\bea\label{eq:Vzcubic}
V_0({\rm cubic}) &=& \lambda_1 v_1\rho_1^3 +\lambda_2 v_2\rho_2^3 +
\thalf\lambda_{345}\left(v_1\rho_1\rho_2^2 + v_2\rho_2\rho_1^2\right)\nn\\
&=& -\lambda_{345}v\left(H(H')^2 + (H')^3\cot 2\beta\right)
-\frac{\Delta\widehat t_1}{64\pi^2\cbeta^2}\left[v\cbeta\left(H\cbeta
-H'\sbeta\right)^3\right] \nn\\
&& -\frac{\Delta\widehat t_2}{64\pi^2\sbeta^2}\left[v\sbeta\left(H\sbeta
+H'\cbeta\right)^3\right].
\eea
Our convention for the triple and quartic couplings of $H_1$, for example, is
that they are the coefficients of $H_1^3$ and $H_1^4$ in these two types of
corrections. Then, the corrections to the triple-Higgs couplings from $V_0$
are:\footnote{The corrections to the $\lambda_{345}$ terms involve
  $\cos\delta$ and $\sin\delta$.  We do not include $\delta$-dependence in
  the $\Delta\widehat t_i$ terms because that would be a two-loop
  correction. Because $\delta$ and $\delta/\beta$ are at most a few
  percent~\cite{Lane:2018ycs}, the effect of including them in these terms is
  negligibly small anyway.}
\bea
\label{eq:lam0111}
\lambda^{(0)}_{H_1H_1H_1} &=& -\lambda_{345}\,v\,\sdelta^2(\cdelta - \sdelta\cot
2\beta) -\frac{(\Delta\widehat t_1\cbeta^2 + \Delta\widehat t_2\sbeta^2)\,v}
{64\pi^2},\\
\label{eq:lam0112}
\lambda^{(0)}_{H_1H_1H_2} &=& +\lambda_{345\,}v\,\sdelta(2\cdelta^2 -
 3\sdelta\cdelta\cot 2\beta - \sdelta^2) +
\frac{3(\Delta\widehat t_1 - \Delta\widehat t_2)\,v\sbeta\cbeta}
{64\pi^2},\qquad\\
\label{eq:lam0122}
\lambda^{(0)}_{H_1H_2H_2} &=& -\lambda_{345}\,v\,\cdelta(\cdelta^2 -
3\cdelta\sdelta\cot 2\beta - 2\sdelta^2)
-\frac{3(\Delta\widehat t_1\sbeta^2 + \Delta\widehat t_2\cbeta^2)\,v}
{64\pi^2}.
\eea

\noindent (ii) To calculate the contributions to the triple-Higgs couplings
from $V_1$, it is appropriate that we use the zeroth-order fields $H$ and
$H'$. Then,
\bea
\label{eq:dV111}
\lambda^{(1)}_{H_1H_1H_1} &=& \left.\frac{1}{6}\,\frac{\partial^3 V_1}
    {\partial H^3}\right\vert_{\langle\,\rangle},\\
\label{eq:dV112}
\lambda^{(1)}_{H_1H_1H_2} &=& \left.\frac{1}{2}\,\frac{\partial^3 V_1}{\partial
    H^2 \,\partial H'}\right\vert_{\langle\,\rangle},\\
\label{eq:dV122}
\lambda^{(1)}_{H_1H_2H_2} &=& \left.\frac{1}{2}\,\frac{\partial^3 V_1}{\partial
    H \,(\partial H')^2}\right\vert_{\langle\,\rangle},
\eea
where, again, $\langle\,\rangle$ means that the derivatives are evaluated at
the vacuum expectation values of the fields. Write $V_1$ as
\be\label{eq:Vonebetax}
V_1 = \frac{1}{64\pi^2} \sum_X \alpha_X M_X^4\left(\beta_X+
  \ln(M_X^2/\LGW^2)\right),
\ee
where $M_X^2 = g_X^2(\Phi_1^\dagg \Phi_1 + \Phi_2^\dagg \Phi_2)$, except that
$m_t^2 = \Gamma_t^2\,\Phi_1^\dagg \Phi_1$ with
$\Gamma_t =\sqrt{2}m_t/v\cos\beta$. This $m_t^2$ affects
$\lambda^{(1)}_{H_1H_1H_2}$ and $\lambda^{(1)}_{H_1H_2H_2}$. The constants
$\alpha_X$, $\beta_X$ and $g_X^2$ can be read off from
Eqs.~(\ref{eq:Vone},\ref{eq:bgmasses}). Then we obtain:
\bea
\label{eq:lam1111}
\lambda^{(1)}_{H_1H_1H_1} &=& \frac{v}{16\pi^2}\sum_X\alpha_X
\left[\frac{M_X^4}{v^4}\left(\beta_X + \frac{13}{6} +
    \ln{\frac{M_X^2}{\LGW^2}}\right)\right]_{\langle\,\rangle}, \\
\label{eq:lam1112}
\lambda^{(1)}_{H_1H_1H_2} &=& \frac{3v\tan\beta}{16\pi^2}\left[\frac{m_t^4}{v^4}
\left(\frac{5}{2} +\ln\frac{m_t^2}{\LGW^2}\right)\right]_{\langle\,\rangle},\\
\label{eq:lam1122}
\lambda^{(1)}_{H_1H_2H_2} &=& \frac{v}{16\pi^2}\sum_{X\neq t}\alpha_X
\left[\frac{M_X^4}{v^4}\left(\beta_X + \frac{3}{2} +
    \ln{\frac{M_X^2}{\LGW^2}}\right)\right]_{\langle\,\rangle} \nn\\
 &&-\frac{3v\tan^2\beta}{16\pi^2}\left[\frac{m_t^4}{v^4}
\left(\frac{5}{2} +\ln\frac{m_t^2}{\LGW^2}\right)\right]_{\langle\,\rangle}.
\eea

The $V_0$ and $V_1$ contributions to the four-Higgs coupling
$\lambda_{H_1H_1H_1H_1}$ are
\bea
\label{eq:HHHHV0}
&&\lambda^{(0)}_{H_1H_1H_1H_1} = -\thalf\lambda_{345}\sdelta^2\left(1 +
  \sdelta\cot 2\beta\right)^2 \nn\\
&&\hspace{3cm} -\frac{(\Delta\widehat t_1\cbeta^2 +\Delta\widehat
  t_2\sbeta^2)}{256\pi^2},  \\
\label{eq:HHHHV1}
&&\lambda^{(1)}_{H_1H_1H_1H_1} = \frac{1}{64\pi^2}\sum_X\alpha_X
 \biggl[\frac{M_X^4}{v^4}\biggl(\beta_X + \frac{25}{6}
     + \ln\frac{M_X^2}{\LGW^2}\biggr)\biggr]_{\langle\,\rangle}.
\eea

\begin{figure}[h!]
\includegraphics[width=1.0\textwidth]{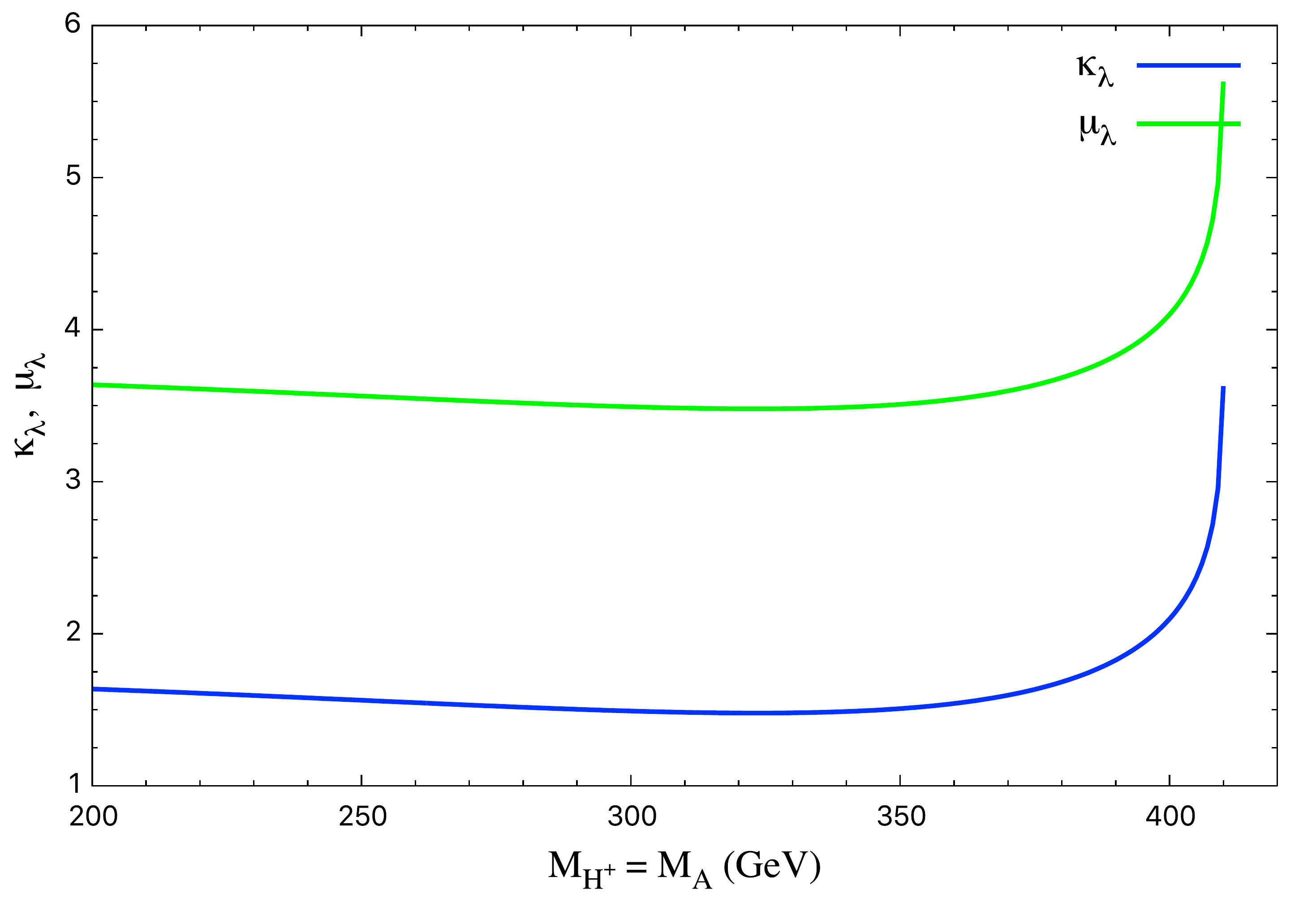}
\caption{The ratios
  $\kappa_\lambda = \lambda^{(0)+(1)}_{H_1H_1H_1}/(\lambda_{HHH})_{\rm SM}$ (with
  $(\lambda_{HHH})_{\rm SM} \cong 32\,\gev$) and
  $\mu_\lambda = \lambda^{(0)+(1)}_{H_1H_1H_1H_1}/(\lambda_{HHHH})_{\rm SM}$
  (with $(\lambda_{HHHH})_{\rm SM} \cong 0.0323$) as a function of
  $M_{H^\pm} = M_A$. The sharp rise starting near $M_{H^\pm} = 370\,\gev$ is
  an artifact of $M_{H'}$ starting its dive to zero.}
\label{fig:kappamu}
\end{figure}

In Fig.~\ref{fig:kappamu} we plot the allowed range of
$\kappa_\lambda = \lambda^{(0)+(1)}_{H_1H_1H_1}/(\lambda_{HHH})_{\rm SM}$ and
$\mu_\lambda = \lambda^{(0)+(1)}_{H_1H_1H_1H_1}/(\lambda_{HHHH})_{\rm SM}$
for this GW-2HDM (where $(\lambda_{HHH})_{\rm SM} = M_H^2/2v \cong 32\,\gev$
and $(\lambda_{HHHH})_{\rm SM} = M_H^2/8v^2 \cong 0.0323$). For this, we put
$M_{H^\pm} = M_A$ to eliminate the scalars' contributions to the
$T$-parameter and then enforced the sum rule~(\ref{eq:MHsum}) so that
$M_{H'} = [(540\,\gev)^4 - 2M_{H^\pm}^4 - M_A^4]^{1/4}$. (We also set
$\tan\beta = 0.5$, its current experimental upper
limit~\cite{Lane:2018ycs}. There is no discernible effect on the cubic and
quartic Higgs couplings for any plausible $\tan\beta > 0$.) From this plot,
we see that $\kappa_\lambda \cong 1.6$ and $\mu_\lambda \cong 3.6$ below
$M_{H^\pm} \simeq 370\,\gev$. In this region, only 2--10\% of these cubic and
quartic Higgs couplings comes from $V_0$. Above it, these couplings
approximately double as the sum rule forces $M_{H'}$ rapidly to zero at
$M_{H^\pm} \cong 410\,\gev$; see Fig~\ref{fig:CPeven}. This is an artifact of
the end point of the sum rule, with the sudden increase due entirely to the
$\Delta\widehat t_{1,2}$ terms in
Eqs.(\ref{eq:lam0111},\ref{eq:HHHHV0}).\footnote{Using $M_{H_2}$ instead of
  $M_{H'}$ from the sum rule lessens somewhat this sharp rise in
  $\kappa_\lambda$ and $\mu_\lambda$, but that is not consistent
  loop-perturbation theory.}

This effect of the sum rule is seen numerically in Table~\ref{tab:lambdas}
where we list triple and quartic couplings for three extreme values of
$M_{H^\pm}$ in Fig.~\ref{fig:kappamu}. The $V_0$~contribution to
$\lambda_{H_1H_1H_2}$ is small and its $V_1$~contribution would vanish were
it not for the fact that $m_t^2 = \Gamma_t \Phi_1^\dagg \Phi_1$ contains a
linear term in~$H'$. On the other hand, for almost the entire $M_{H^\pm}$
range, the contributions to $\lambda_{H_1H_2H_2}$ listed in the table are of
normal size, $\CO(\lambda_{HH'H'} = M_{H_2}^2/v$). The interesting question
of the effect this large coupling has on the production rate of
$pp \to H_2 H_2$ is beyond the scope of this paper.

\begin{table}[!h]
     \begin{center}{
  \begin{tabular}{|c|c|c|c|c|}
  \hline
$M_H \cong M_{H_1}$ & $M_{H^\pm} = M_A$ & $M_{H'}$ ($M_{H_2}$) &
$\lambda^{(0)+(1)}_{H_1H_1H_1}$& $\kappa_{\lambda}$\\
  \hline
125  & 200 & 532 (534) & 51.9 & 1.64 \\
125  & 400 & 301 (314) & 66.6 & 2.10 \\
125  & 410 & 115 (214) & 115  & 3.63  \\
\hline\hline
$\lambda^{(0)+(1)}_{H_1H_1H_2}$& $\lambda_{HH'H'}$&$\lambda^{(0)+(1)}_{H_1H_2H_2}$
&$\lambda^{(0)+(1)}_{H_1H_2H_2}/\lambda_{HH'H'}$ &\\
\hline
3.84 & 1151 & 1252 & 1.09 &\\
2.62 & 367  & 510  & 1.39 &\\
7.92 & 54.1 & 349  & 6.46 &\\
\hline\hline
$\lambda^{(0)}_{H_1H_1H_1H_1}$& $\lambda^{(1)}_{H_1H_1H_1H_1}$
& $\lambda^{(0)+(1)}_{H_1H_1H_1H_1}$& $\mu_\lambda$ & \\
\hline
$-0.942\times 10^{-3}$ & 0.118& 0.117& 3.64 & \\
0.0139& 0.118& 0.132& 4.10 & \\
0.0634& 0.118& 0.182& 5.63 & \\
\hline
  \end{tabular}}
\caption{Selected cubic and quartic couplings of the 125~GeV Higgs
  boson. Input masses are $M_H \cong M_{H_1}$ and $M_{H^\pm} = M_A$, with
  $M_{H'}$ taken from the sum rule Eq.~(\ref{eq:MHsum}) as explained in the
  text; $M_{H_2}$ is the corresponding $C\!P$-even eigenvalue at one-loop order;
  $\tan\beta = 0.50$, and the misalignment angle $\delta = 0.0039$, 0.0115,
  0.0323 for $M_{H^\pm} = 200$, 400, $410\,\gev$. Couplings $\lambda^{(0)}$
  and $\lambda^{(1)}$ are contributions from the one-loop improved $V_0$ and
  full one-loop $V_1$ potentials. Comparisons are made to the Standard Model
  ($\kappa_\lambda = \lambda^{(0)+(1)}_{H_1H_1H_1}/(\lambda_{HHH})_{\rm SM}$
  and
  $\mu_\lambda = \lambda^{(0)+(1)}_{H_1H_1H_1H_1}/(\lambda_{HHHH})_{\rm SM}$)
  or to tree-level values ($\lambda^{(0)+(1)}_{H_1H_2H_2}/\lambda_{HH'H'}$).
  Masses and cubic couplings are in GeV units.}
\label{tab:lambdas}
\end{center}
\end{table}

The value of the triple-Higgs coupling $\lambda^{(0)+(1)}_{H_1H_1H_1}$
in the GW-2HDM is close to its small SM value. As can be seen in
Refs.~\cite{Sirunyan:2018two, Aad:2019uzh,Carvalho:2016rys},
$\kappa_\lambda \simeq 1.6$--$3.6$ corresponds to
$\sigma(pp \to HH) = 15$--$20\,\fb$. This is the absolute minimum value of
the di-Higgs production cross section for $\sqrt{s} = 13$--$14\,\tev$ at the
LHC. Because the sum rule~(\ref{eq:gensum}) is independent of the number or
type of Higgs multiplets in the GW model, this result is true of
{\em all} of them.


We are aware that there are many theoretical studies of the cubic and even
quartic Higgs couplings --- in the context of one-doublet models,
multi-doublet models, models with extra singlet ``Higgses'', and so on ---
many more studies than we can note here. We apologize to their authors for
not citing them. At perhaps the simplest level, this is the problem of the
shape of the potential of {\em the} Higgs boson itself, specifically, what
are $\lambda_{HHH}$ and $\lambda_{HHHH}$? One recent
paper~\cite{Agrawal:2019bpm} studied a variety of new physics scenarios,
their effect on these couplings, and the prospect of distinguishing them at
the $14\,\tev$ High Luminosity LHC (HL-LHC), the $27\,\tev$ High Energy LHC
(HE-LHC) and the $100\,\tev$ Future Circular Hadron Collider (FCC-hh). These
authors considered, {\em inter alia}, a Coleman-Weinberg-like
potential. Compared to the SM values, they found $\kappa_\lambda = 5/3$ and
$\mu_\lambda = 11/3$. These are close to our calculated values of
$\kappa_\lambda$ and $\mu_\lambda$ in Fig.~\ref{fig:kappamu} below
$M_{H^\pm} = M_A \cong 370\,\gev$. According to the analysis in
Ref.~\cite{Agrawal:2019bpm} of di-Higgs and tri-Higgs observability at the
upgraded LHC and the FCC-hh, the HE-LHC is needed to detect and distinguish
the triple Higgs coupling of the GW-2HDM and the FCC-hh is needed for the
quartic coupling. This is a gloomy prospect.


\section*{IV. Testing Gildener-Weinberg at the LHC}

\begin{figure}[h!]
\includegraphics[width=1.0\textwidth]{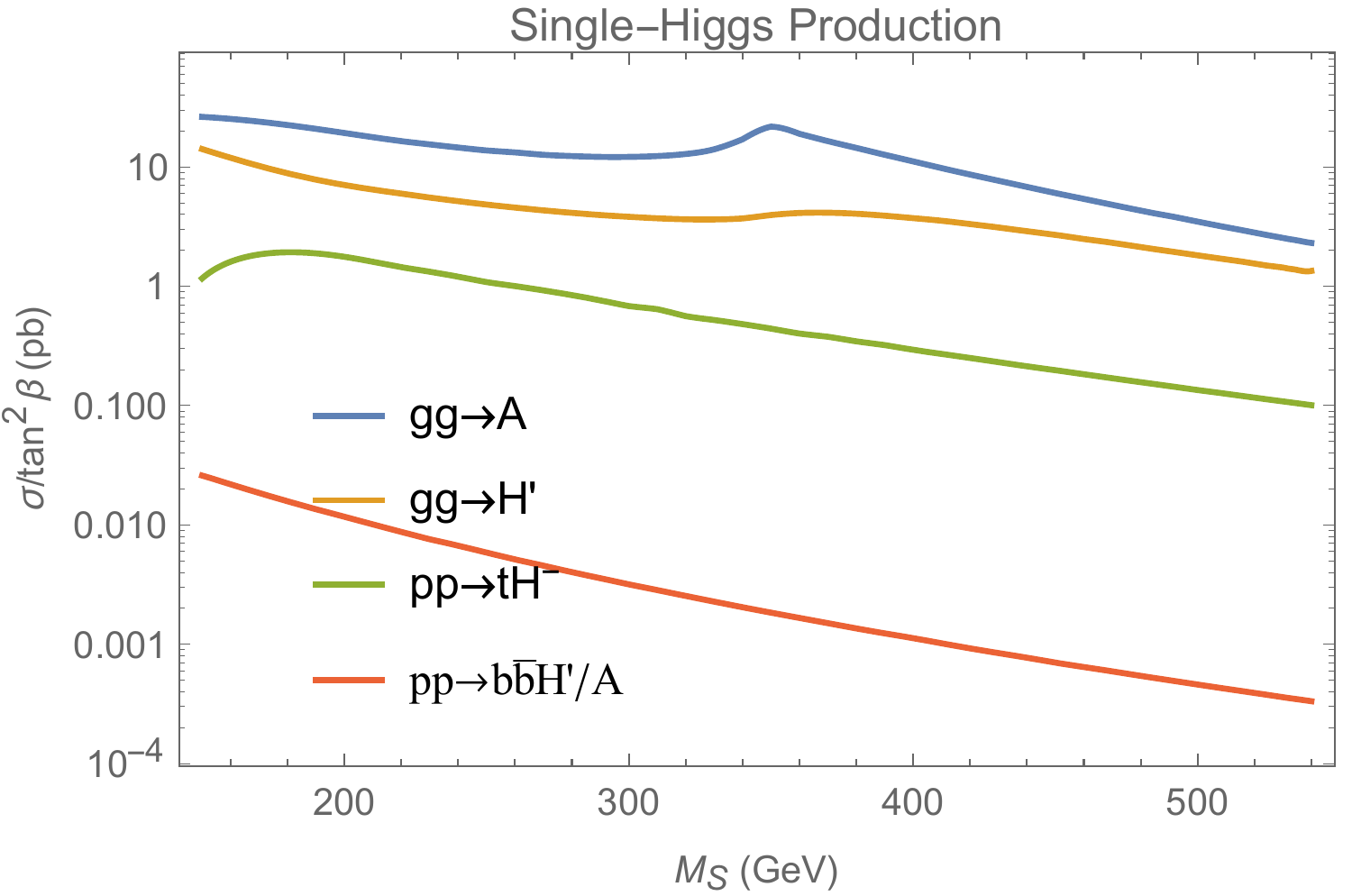}
\caption{The cross sections for $\sqrt{s} = 13\,\tev$ at the LHC for single
  Higgs production processes in the alignment limit ($\delta \to 0$) of the
  GW-2HDM with the dependence on $\tan\beta$ scaled out. Both charged Higgs
  states are included in $pp \to t H^-$. From Ref.~\cite{Lane:2018ycs}.}
\label{fig:sigmas}
\end{figure}

Much more immediately promising avenues of attack on GW models are searches
for the new charged and neutral Higgs bosons that lie below
400--$500\,\gev$. In the GW-2HDM, the new scalars are just $H^\pm$, $A$ and
$H_2$. Assuming as we have that $M_{H^\pm} = M_A$, the principal search modes
are:
\bea\label{eq:Hsearches}
H^\pm &\to& t\bar b\,(b\bar t)\,\,{\rm and}\,\,W^\pm H_2;\\
A &\to& b\bar b,\,\,t\bar t \,\,{\rm and}\,\,ZH_2;\\
H_2 &\to& b\bar b,\,\,t\bar t \,\,{\rm and}\,\,ZA,\,W^\pm H^\mp.
\eea

Their main production cross sections at the 13~TeV LHC were discussed in
Ref.~\cite{Lane:2018ycs} and they are displayed in Fig.~\ref{fig:sigmas} with
the dependence on $\tan^2\beta$ scaled out. There seem to have been but a few
searches for $pp \to H_2\,\,{\rm or}\,\, A \to b \bar b$, presumably because
of the overwhelming continuum $b\bar b$ production. One recent search by CMS
for a $C\!P$-even or odd scalar with $M_{b\bar b} = 50$--$350\,\gev$ and produced
at high-$p_T$ is reported in Ref.~\cite{Sirunyan:2018ikr}. No significant
excess over SM backgrounds was found. For $M_{H_2,A} = 200$--$350\,\gev$, the
95\%~CL limits are
$\sigma(pp \to H_2,A)B(H_2,A \to b \bar b) \simeq 200$--$300\,\pb$ which
translates into upper limits $\tan\beta \simeq 3$--6. It is important to note
that the decays $H_2,A \to W^+W^-,\,\,ZZ$ and $H^\pm \to W^\pm Z$ are highly
suppressed in GW models by the near alignment of the SM Higgs $H \cong H_1$.
Likewise, alignment strongly suppresses $H_2,A \to ZH$ and
$H^\pm \to W^\pm H$. Seeing these decay modes from a new, heavier
{\em spinless} boson would be significant, if not fatal, blows to GW models.

The following is a summary of the current experimental situation for the
new Higgs bosons' dominant decay modes:

\begin{itemize}

\item[(1)] The CMS search at $8\,\tev$ for
  $H^\pm \to t\bar b$~\cite{Khachatryan:2015qxa} restricted
  $\tan\beta \simle 0.5$ for our type-I GW-2HDM with
  $180\,\gev < M_{H^\pm} < 550\,\gev$~\cite{Lane:2018ycs}. Searches at
  $13\,\tev$ for $H^\pm \to t\bar b$ by ATLAS~\cite{Aaboud:2018cwk} and
  CMS~\cite{Sirunyan:2019arl} extend down to $M_{H^\pm} = 200\,\gev$, but do
  not yet have the sensitivity to reach $\sigma(pp \to t H^\pm) = 0.50\,\pb$
  ($0.033\,\pb$) expected at $200\,\gev$ ($500\,\gev$) for $\tan\beta = 0.50$
  and $B(H^\pm \to t\bar b) = 1$. At $M_{H^\pm} = 400\,\gev$ in the GW-2HDM,
  Fig.~\ref{fig:CPeven} gives $M_{H_2} = 314\,\gev$, while at
  $M_{H^\pm} = 408\,\gev$, it gives $M_{H_2} = 247\,\gev$. Between these two
  mass points, the $H^\pm \to W^\pm H_2$ decay rate increases by a factor
  of~70, overwhelming the $H^\pm \to t\bar b$ decay rate; see item~(3)
  below. The two processes $g \bar b \to H^+ \bar t, \,H^+ \to t \bar b$ and
  $g \bar b \to H^+ \bar t, \,H^+ \to W^+ H_2$, with $H_2 \to b\bar b$, have
  the same final state. Hence, $H^+ \to W^+ H_2$ may unintentionally be
  included in a search for $H^+ \to t \bar b$.  Even if that happened, the
  model expectation $\sigma(g\bar b \to H^+ \bar t) = 0.075\,\pb$ for
  $M_{H^\pm} \simeq 400\,\gev$ and $\tan\beta = 0.5$ is well below the 95\%~CL
  limits $\simeq 0.5\,\pb$ (ATLAS) and $0.7\,\pb$ (CMS). There appear to be
  no dedicated searches released for
  $H^\pm \to W^\pm H_2 \to \ell^\pm b \bar b$ and for
  $H_2 \to W^\pm H^\mp \to \ell^\pm t \bar b$.

\begin{table}[!t]
     \begin{center}{
  \begin{tabular}{|c|c|c|c|c|}
  \hline
$M_A = M_{H^\pm}$ & $M_{H_2}$ & ATLAS & CMS & GW-2HDM\\
\hline
400 & 300 & 255 &  75 &  65\\
300 & 500 & 105 &  50 & 100\\
\hline
\end{tabular}}
\caption{95\% CL upper limits on
  $\sigma(pp \to A(H_2))\, B(A(H_2) \to ZH_2(A)) B(H_2(A) \to \bar bb)$ via
  gluon fusion from ATLAS~\cite{Aaboud:2018eoy}, CMS~\cite{Sirunyan:2019wrn}
  and GW-2HDM calculations for two cases of large $M_A$ and $M_{H_2}$. The
  CMS limits include $B(Z \to e^+e^-,\,\mu^+\mu^-)$; the ATLAS limits and
  GW-2HDM predictions do not. Masses are in GeV and $\sigma B$
  in~femtobarns. $M_A = M_{H^\pm}$ is assumed and $M_{H_2}$ is taken from
  Fig.~\ref{fig:CPeven} as explained in the text. Model cross sections are
  taken from Fig.~\ref{fig:sigmas} multiplied by $\tan^2\beta = 0.25$.}
\label{tab:limits}
\end{center}
\end{table}

\item[(2)] CMS recently reported a search for a~$C\!P$-even or odd
  scalar~$\varphi$ with mass in the range 400~to $700\,\gev$ and decaying to
  $t \bar t$~\cite{Sirunyan:2019wph}. Results were presented in terms of
  allowed and excluded regions of the ``coupling
  strength''~$g_{\varphi} = \lambda_{\varphi t\bar t}/(m_t/v)$ and for fixed
  width-to-mass ratio $\Gamma_\varphi/M_\varphi = 0.5$--25\%. In the GW-2HDM,
  $g_{\varphi} = \tan\beta$. For the $C\!P$-odd case, $\varphi = A$, with
  $400\,\gev < M_A < 500\,\gev$ and all $\Gamma_A/M_A$ considered, the region
  $\tan\beta < 0.5$ is not excluded.\footnote{The same appears to be true for
    $\varphi = H_2$ with $\Gamma_{H_2}/M_{H_2} \simge 1\%$.} This is possibly
  due to an excess at $400\,\gev$ that corresponds to a global (local)
  significance of $1.9\,\,(3.5\pm 0.3)\,\sigma$ for
  $\Gamma_A/M_A \simeq 4\%$. Ref.~\cite{Sirunyan:2019wph} also notes that
  $t\bar t$ threshold effects may account for the excess.

\item[(3)] Searches for $pp \to A(H_2) \to ZH_2(A) \to \ell^+\ell^- b \bar b$
  via gluon fusion have been reported by ATLAS~\cite{Aaboud:2018eoy} and
  CMS~\cite{Sirunyan:2019wrn}. Two examples of observed 95\% upper limits on
  cross sections and the corresponding GW-2HDM predictions are given in
  Table~\ref{tab:limits}. A word of caution is in order here: These decay
  rates are dominated by the emission of longitudinally-polarized weak bosons
  and are proportional to $p^3/M_{W,Z}^2$, hence sensitive to the available
  phase space.

\end{itemize}

At the LHC there are now $140\,\ifb$ of $pp$~collision data at $13\,\tev$
from Run~2 and another $200\,\ifb$ at $14\,\tev$ are expected from Run~3 by
the time it concludes at the end of 2024. With masses in the range
$200$--$500\,\gev$, GW~Higgs production rates are
$\sigma(pp \to H^+ + H^-) = (0.1-1.0)\,\pb \times \tan^2\beta$,
$\sigma(pp \to A) = (4.0-20)\,\pb \times \tan^2\beta$ and
$\sigma(pp \to H_2) = (2.0-7.0)\,\pb \times \tan^2\beta$. Thus, unless
$\tan\beta \simle 0.2$, there will be anywhere from $10^3$ to several $10^6$
of these GW~Higgs bosons produced by the end of Run~3. Given the large SM
production of $b\bar b$, direct detection of $H_2 \to b \bar b$ via gluon
fusion is the most difficult. There is no doubt that improved sensitivity in
the low-mass region of $H^\pm \to t\bar b$ is needed to access the expected
cross sections. The decays
$A({\rm or}\,\,H_2) \to ZH_2({\rm or}\,\,A) \to \ell^+ \ell^- b \bar b$,
$H^\pm \to W^\pm H_2 \to \ell^\pm b \bar b + \etmiss$ are helped by the
narrow $b \bar b$ resonance and lepton kinematics. They may be easier than
$H^\pm \to t \bar b$, but they cover a slimmer portion of
($M_{H^\pm},M_A, M_{H_2}$)-space, the upper and lower ends of the allowed
$M_{H^\pm} = M_A$ region.

 \section*{Acknowledgments}

 We are grateful for informative conversations with and advice from Tulika
 Bose, Kevin Black, Gustaaf Brooijmans, Jon Butterworth, Estia Eichten,
 Howard Georgi, Guoan Hu, Greg Landsberg, Kimyeong Lee, William Murray,
 Alessia Saggio, David Sperka and Erick Weinberg. KL acknowledges the warm
 hospitality of the CERN Theory Division and Laboratoire d'Annecy-le-Vieux de
 Physique Th\'eorique and valuable interactions at the PhysTeV meeting at Les
 Houches in July~2019.

\vfil\eject

\bibliography{GW_Higgs}
\bibliographystyle{utcaps}
\end{document}